\documentclass[a4paper,
             ]{paper}
\usepackage{siunitx}
\usepackage{booktabs}
\usepackage{textcomp}
\usepackage{graphicx}

\begin{document}

\title{On the suitability of longitudinal profile measurements using Coherent Smith-Purcell radiation for high current proton beams\thanks{Work financially supported by the Universit\'e Paris-Sud (programme "attractivit\'e") and the French ANR (contract ANR-12-JS05-0003-01)}}

\author{J. Barros\thanks{barros@lal.in2p3.fr}, N. Delerue, M. Vieille-Grosjean, LAL, Orsay, France\\
        I. Dolenc Kittelmann, C. Thomas, ESS, Lund, Sweden}

\maketitle

\begin{abstract}
The use of Smith-Purcell radiation to measure electrons longitudinal profiles has been demonstrated at several facilities in the picosecond and sub-picosecond range. There is a strong interest for the development of non intercepting longitudinal profile diagnostics for high current proton beams. We present here results of simulations on the expected yield of longitudinal profile monitors using Smith-Purcell radiation for such proton beams.
\end{abstract}

\section{introduction}

Handling high current proton beams is a challenge for beam diagnostics in future proton accelerators. In order to achieve a high beam power, it is mandatory to understand the beam dynamics and to determine the beam size and profile with precision. In high power proton Linacs, such understanding would allow a full characterization of the beam after each Linac section. With a beam transverse size of 2 to \SI{3}{mm}, intercepting profile diagnostics such as OTR screens or Cherenkov monitors can be used only during tuning and specific beam operation modes. To determine the longitudinal profile, Feschenko-type Bunch Shape Monitors (BSM \cite{Feschenko}) can be used in the low energy sections of the Linac, however, BSM may not be sensitive to short bunch lengths. In the higher energy sections, it is highly advisable to use non interceptive devices to avoid radiations or damage to the equipment. Wall current monitors or other methods based on detecting the fields at the vacuum chamber boundary could be used, but they are intrinsically limited in resolution due to the rather low relativistic $\beta$ \cite{ESS_TDR} and do not have an adapted temporal resolution. Therefore, a diagnostic based on Smith-Purcell radiation could be of high interest. We will investigate here the possibility of using a Smith-Purcell diagnostic for the measurement of proton beams longitudinal profiles, using a numerical simulation code presented hereafter.

\section{description of the simulation}
\subsection{Theoretical considerations}
\begin{figure}
   \centering
   \includegraphics*[width=60mm]{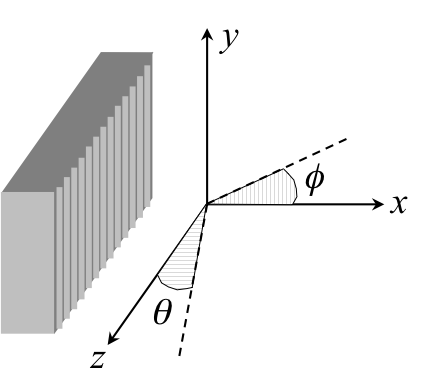}
   \caption{Coordinate system used for the simulation}
   \label{fig:Coordinates}
\end{figure}

Smith-Purcell radiation can be described by the "surface current" theory \cite{Brownell_PRE_1998}: when charged particles travel over a grating, they induce an image charge on its surface and keep pace with it. The corrugations of the grating cause the created current to accelerate, which in turn leads to the emission of radiation. In our coordinate system, we assume that the protons travel perpendicularly to the grooves of a blazed grating. Due to the periodic structure of the grating, the emitted wavelength $\lambda$ depends on the observation angle $\theta$ defined in Fig. \ref{fig:Coordinates}. The grating shape and pitch defines its efficiency, that varies depending on $\theta$ and on the second observation angle called $\phi$.

Our simulation describes the Smith-Purcell intensity radiated per unit solid angle for a single particle, as described in the equation (2a) of \cite{Doucas_PRSTAB_2006} for a grating of period \textit{l} and length Z positioned at a distance $x_0$ from the beam center:

\begin{equation}
\label{eq:singleParticle}
    \left(\frac{dI}{d\Omega}\right)_1=2\pi q^2 \frac{Zl}{n}\frac{1}{\lambda^3}R^2exp\left[ -\frac{4\pi x_0}{\lambda}\frac{\sqrt{1+\beta^2\gamma^2 sin^2 \theta sin^2 \phi}}{\beta\gamma}\right]
\end{equation}

The term $R^2$ represents the grating efficiency that contains the contribution from each period of the grating, and is different for each grating shape \cite{Brownell_PRSTAB_2005}. Here we use echelette gratings whose efficiency will be shown in the Results section.

For a given bunch of $N_P$ particles, the emitted radiation has a high degree of coherence when the radiation wavelength is comparable to the bunch length. In the equation (\ref{equ:coherence}), $S_{coh.}$ and $S_{incoh.}$ represent the coherent and incoherent components of the emitted radiation respectively. Assuming that the transverse and longitudinal profiles are not correlated, $S_{coh.}$ can be expressed as the product two functions \textit{T} and \textit{G} encoding the longitudinal (temporal) and transverse profile respectively. $\sigma_x$ and $\sigma_y$ are the beam dimensions in x and y directions, and $\omega$ is the frequency.

\begin{equation}
\label{equ:coherence}
\left(\frac{dI}{d\Omega}\right)_{N_P}=\left(\frac{dI}{d\Omega}\right)_{1}(N_P S_{incoh.}+N^2_P S_{coh.})\\
\end{equation}

\begin{equation}
\label{equ:Scoh}
S_{coh.} = \left|\int_{-\infty}^{\infty}Te^{-i\omega t}dt\right|^2 G^2(\sigma_x,\sigma_y)\\
\end{equation}

In the case where the coherent component prevails, the emitted radiation encodes the form factor of the bunch and its measurement allows to recover information about the bunch longitudinal profile.
If we assume that the beam is gaussian in the transverse plane and \textit{G} is normalized to 1, then the emitted energy is proportional to the form factor as shown in \ref{equ:approxCoherence}.

\begin{equation}
\label{equ:approxCoherence}
\left(\frac{dI}{d\Omega}\right)_{N_P}\simeq\left(\frac{dI}{d\Omega}\right)_{1}N^2_P\left|\int_{-\infty}^{\infty}Te^{-i\omega t}dt\right|^2
\end{equation}

This form factor corresponds to the modulus of the Fourier transform of the temporal profile. In order to recover the full profile the phase needs to be reconstructed. Several techniques are possible, and are documented in \cite{Blackmore_2008}.
 
In this work, we use a code package written by G. Doucas \cite{Doucas_PRSTAB_2006}, that takes several beam parameters as an input and generates the a Smith-Purcell spectrum, which is a calculation made from the Fourier transform of the temporal profile and chosen grating parameters.

\subsection{Beam Parameters}

The parameters used for this study on proton beams are summarized in Table \ref{tab:BeamParameters}. We assume that the beam transverse dimensions would allow approaching a grating at a distance $x_0=\SI{10}{mm}$ from the beam center.

\begin{table}[hbt]
   \centering
   \caption{Used beam parameters for the simulation}
   \begin{tabular}{lcc}
       \toprule
       \textbf{Parameter} & \textbf{Unit} & \textbf{Value}\\
       \midrule
           $\gamma$         & - & $\approx$3                    \\
           $\beta$         & - & 0.92                    \\
          Transverse x size (FWHM)       & mm & 2                  \\
          Transverse y size (FWHM)       & mm & 2                  \\
          Number of protons per bunch       & - & \num{1e9}    \\
          Expected bunch length (FWHM)      & ps & 3            \\
          Number of bunches per train         & - & \num{1e6}	\\
          Frequency				& MHz & 352		\\
          Repetition rate				& Hz & 14			\\
       \bottomrule
   \end{tabular}
      \label{tab:BeamParameters}
\end{table}

\section{choice of measurement parameters}

\subsection{Detection Parameters}
The choice of suitable grating parameters is made by first taking into account the preferred geometry for detection. Here we chose to maximize the signal at $\theta$=90\textdegree.  Our optics collect light with a high numerical aperture, so that the signal generated by the code will be integrated over the whole captured solid angle. The optics are placed at \SI{200}{mm} from the grating, and the opening of the detection cone is \SI{100}{mm} to increase the incoming signal on the detectors.

\subsection{Grating Parameters}
\begin{figure}[hbt]
   \centering
   \includegraphics*[width=65mm]{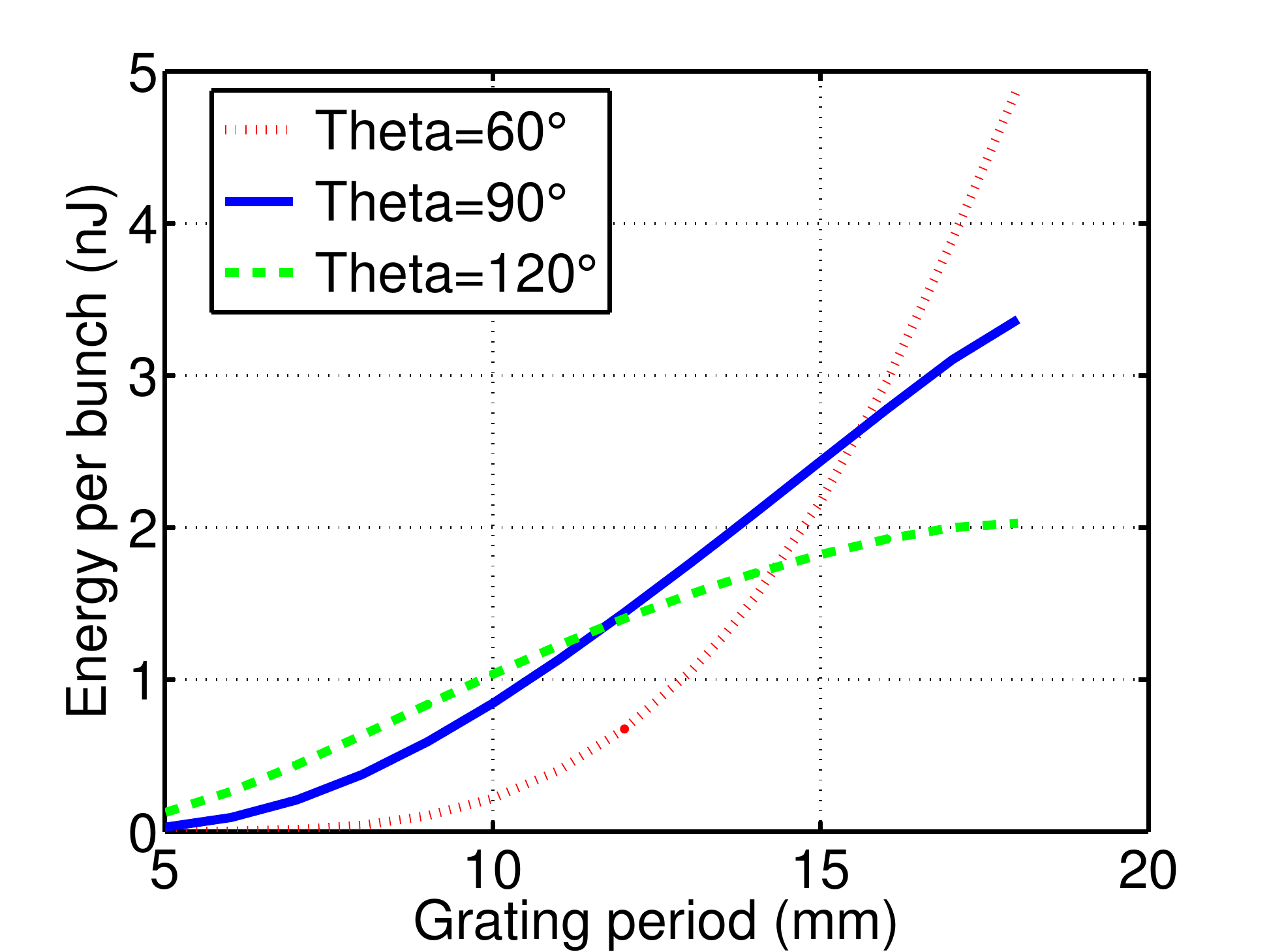}
   \caption{Evolution of the Smith-Purcell signal with different grating pitches.}
   \label{fig:GratingPitch}
\end{figure}

The grating pitch is selected in order to make sure that the emission is mostly in the coherent regime, so that the measurement would be sensitive to the beam longitudinal profile (see Eq. (\ref{equ:coherence})). The increase in signal seen on Fig. \ref{fig:GratingPitch} when the grating pitch increases, corresponds to the onset of the coherent component. We chose a value of the grating period that is a trade-off between the different angles to ensure a good sensitivity at all angles: the three angular components are in the coherent range. Keeping in mind that the insertion length needs to be as small as possible, and that we would prefer the maximum of the signal to be at 90\textdegree, we chose a grating period of \SI{13}{mm}. The grating parameters are summarized in Table \ref{tab:GratingParameters}.

\begin{table}[hbt]
   \centering
   \caption{Grating parameters for the simulation}
   \begin{tabular}{lcc}
       \toprule
       \textbf{Parameter} & \textbf{Value}\\
       \midrule
          Width      & \SI{40}{mm}                      \\
          Number of periods       &15                    \\
          Blaze angle       & 30\textdegree          \\
          Pitch &\SI{13}{mm}\\
       \bottomrule
   \end{tabular}
      \label{tab:GratingParameters}
\end{table}

\section{Results}
\begin{figure}[hbt]
   \centering
   \includegraphics*[width=65mm]{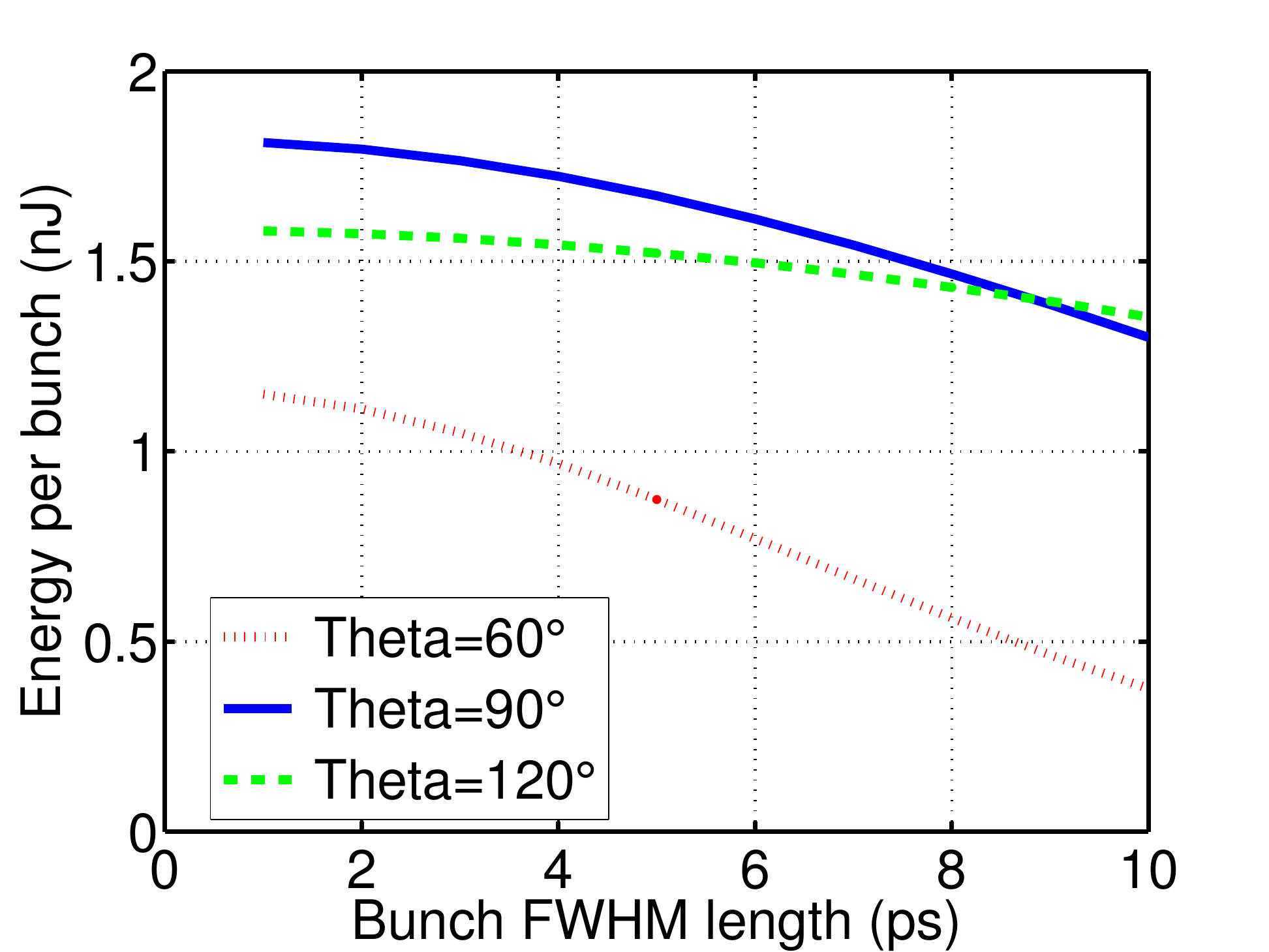}
   \caption{Simulated Smith-Purcell signal for a 13 mm grating depending on the bunch length.}
   \label{fig:BunchLength}
\end{figure}

Knowing the expected signal level, it is now necessary to verify that the device is sensitive to a longitudinal profile change, which is a crucial feature for a profile measurement tool. The sensitivity to a variation in the bunch FWHM length is shown in Fig. \ref{fig:BunchLength} for the chosen grating and optics parameters. The coherent emission becomes progressively predominant as the bunch length decreases, which leads to a significant and measurable signal increase. This is shown in Fig. \ref{fig:BunchLength}, where the wavelength at $\theta$ = 90\textdegree varies by more than 20\% for a bunch length varying from 3 to \SI{10}{ps}, but a similar change is also observed at other wavelengths showing a significant change of the power spectrum distribution. 

\begin{figure}[hbt]
   \centering
   \includegraphics*[width=70mm]{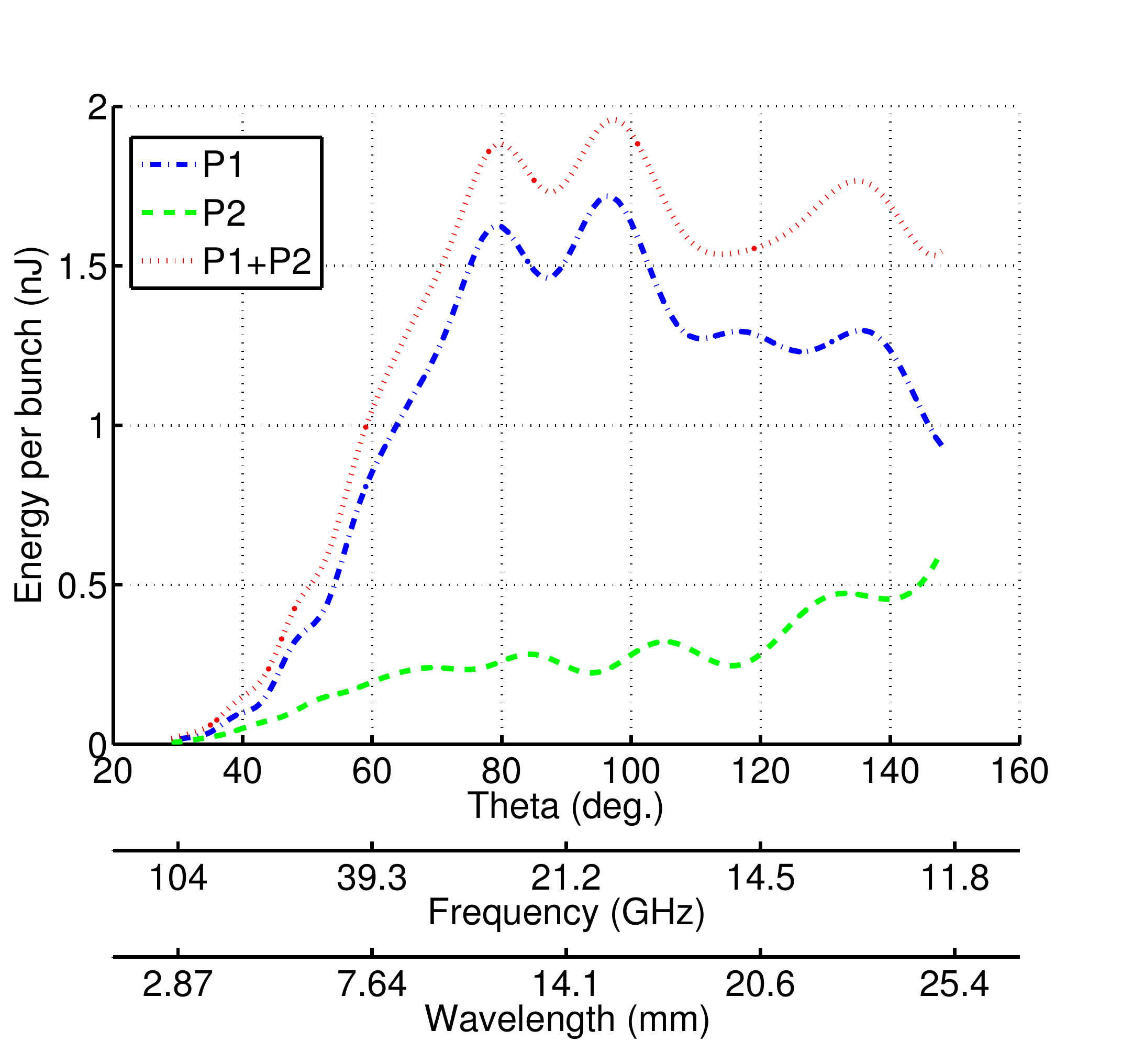}
   \caption{Simulation of the expected signal for a 3 ps bunch. P1 and P2 are the two polarization components.}
   \label{fig:Theta}
\end{figure}

The signal level for the simulated case, as shown in Fig. \ref{fig:BunchLength}, is of the order of \SI{1}{nJ} per bunch, and needs to be compared with the performances of the detectors used. For example, if the detectors measure an averaged signal over several ms, which is the case of the pyroelectric sensors often used in experiments involving electrons \cite{Andrews_PRSTAB_2014}, they will detect the Smith-Purcell power produced by a whole train (around \num{1e6} bunches), thus a higher signal by several orders of magnitude, up to mJ. However, these values do not take into account the transmission of the different optical elements that would need to be installed in front of the detectors. Also, as seen in Fig. \ref{fig:Theta}, the expected Smith-Purcell emission would occur in the millimeter-wave range where the background level needs to be evaluated to be able to distinguish it from the actual signal. If we use detectors that are sensitive only to signal variations, they will not impacted by the effect of the temperature. However, electromagnetic noise and other sources of radiations related to the beam can create background. Since the Smith-Purcell radiation is linearly polarized, a way to decorrelate the signal and the background could then be to measure the two polarization components.

\begin{figure}[htb]
   \centering
   \includegraphics*[width=90mm]{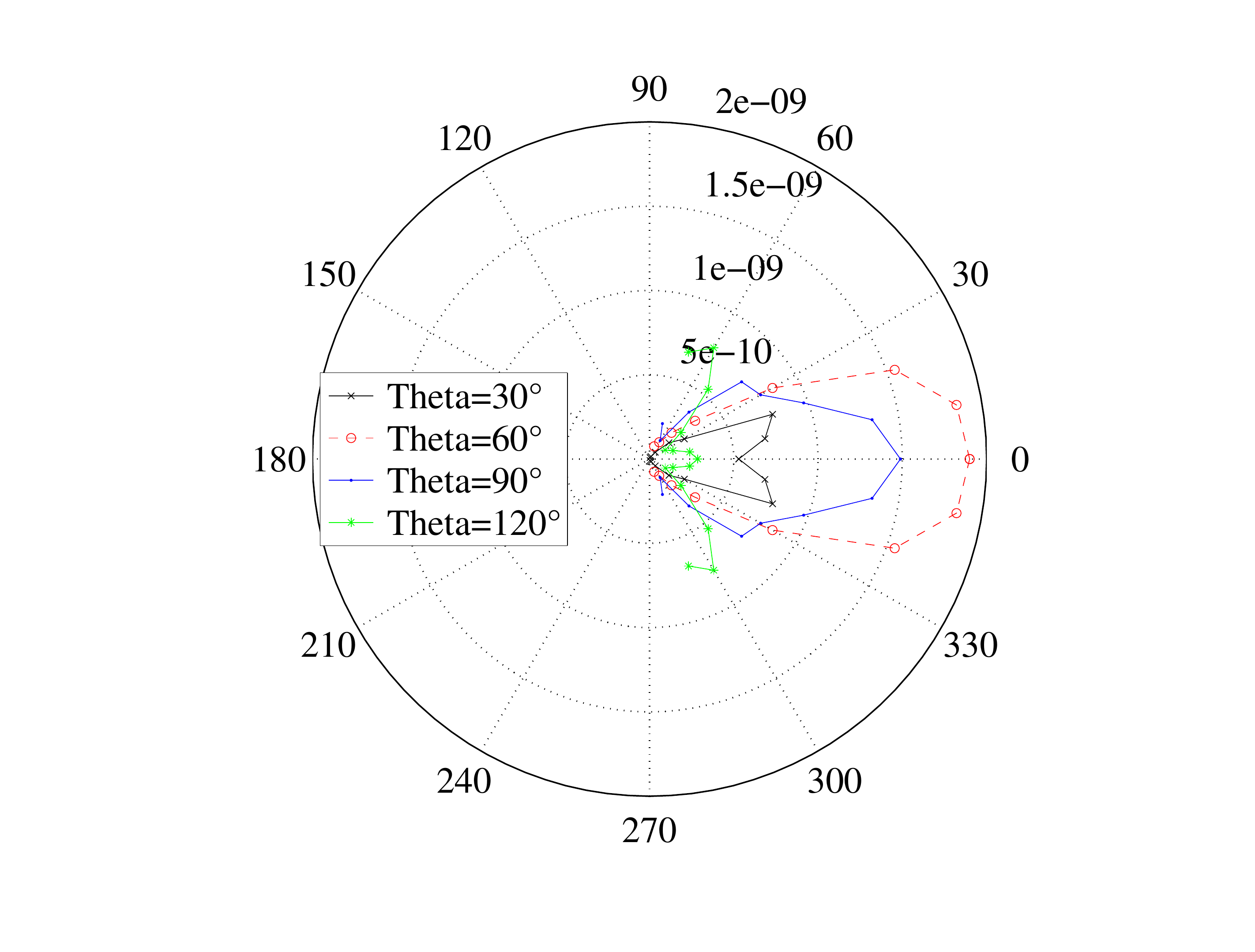}
   \caption{Efficiency of the grating depending on the $\phi$ angle, for different $\theta$ values. The emission source is the center of the figure, the detection optics would be placed on the right half of the figure.}
   \label{fig:Phi}
\end{figure}

The efficiency of the grating is given in Fig. \ref{fig:Phi} depending on the $\phi$ angle. This indicates that most of the radiation would be emitted along the x axis in a cone of approximately 40\textdegree opening, allowing all signal to enter the \SI{100}{mm} optics.

\section{Conclusion}
Our simulations indicate that with the parameters used for our simulation a longitudinal bunch profile monitor based on Coherent Smith-Purcell radiation installed at near a high intensity proton beam would be sensitive to changes in the pulse length and could be useful in the tuning phase of such accelerator. However such device has never been tested at low $\beta$ proton beams and preliminary tests on a high intensity proton source would be useful.

\section{Acknowledgements}
The authors acknowledge George Doucas from John Adams Institute in Oxford for his useful advice.

\end{document}